\documentclass[a4paper,conference]{IEEEtran}
\IEEEoverridecommandlockouts
\usepackage{cite}
\usepackage{amsmath,amssymb,amsfonts}
\usepackage{algorithmic}
\usepackage{graphicx}
\usepackage{textcomp}
\usepackage{xcolor}
\usepackage{float}
\usepackage{afterpage}
\usepackage{enumitem}
\newlist{steps}{enumerate}{1}
\setlist[steps, 1]{label = Step \arabic*:}
\usepackage{graphicx} 
\usepackage{multirow} 
\usepackage{bm} 
\usepackage{stfloats}
\def\BibTeX{{\rm B\kern-.05em{\sc i\kern-.025em b}\kern-.08em
    T\kern-.1667em\lower.7ex\hbox{E}\kern-.125emX}}
\begin{document}

\IEEEoverridecommandlockouts

\title{Optimizing VO\textsubscript{2}max Prediction in Gamified Cardiac Assessment: Leveraging Effective Feature Selection and Refined Protocols for Robust Models\\
{\footnotesize \textsuperscript{}}
\thanks
}
\makeatletter
\newcommand{\linebreakand}{%
  \end{@IEEEauthorhalign}
  \hfill\mbox{}\par
  \mbox{}\hfill\begin{@IEEEauthorhalign}
}
\makeatother
\author{\IEEEauthorblockN{Vaishnavi C K}
\IEEEauthorblockA{\textit{Department of Electrical Engineering} \\
\textit{Indian Institute of Technology Madras}\\
Chennai, India \\
ee23s028@smail.iitm.ac.in}
\and
\IEEEauthorblockN{Sricharan V}
\IEEEauthorblockA{\textit{Department of Electrical Engineering} \\
\textit{Indian Institute of Technology Madras}\\
Chennai, India \\
ee21s068@smail.iitm.ac.in}
\and
\IEEEauthorblockN{Sri Gayathri G}
\IEEEauthorblockA{\textit{Department of Electrical Engineering} \\
\textit{Indian Institute of Technology Madras}\\
Chennai, India\\
srigayathri@htic.iitm.ac.in}
\and
\linebreakand
\IEEEauthorblockN{Danush Adhithya N}
\IEEEauthorblockA{\textit{Department of Electrical Engineering} \\
\textit{Indian Institute of Technology Madras}\\
Chennai, India\\
danush@htic.iitm.ac.in}
\and
\IEEEauthorblockN{Alex Joseph}
\IEEEauthorblockA{\textit{Department of Electrical Engineering} \\
\textit{Indian Institute of Technology Madras}\\
Chennai, India \\
alexsanjoseph@gmail.com}
\and
\IEEEauthorblockN{Preejith SP}
\IEEEauthorblockA{\textit{Department of Electrical Engineering} \\
\textit{Indian Institute of Technology Madras}\\
Chennai, India \\
preejithsp@gmail.com}
\and
\linebreakand
\IEEEauthorblockN{Mohanasankar Sivaprakasam}
\IEEEauthorblockA{\textit{Department of Electrical Engineering} \\
\textit{Indian Institute of Technology Madras}\\
Chennai, India \\
mohan@ee.iitm.ac.in}
}

\maketitle

\begin{abstract}
VO\textsubscript{2}max is a critical indicator of cardiopulmonary fitness, reflecting the maximum amount of oxygen the body can utilize during intense exercise. Accurately measuring VO\textsubscript{2}max is essential for assessing cardiovascular health and predicting outcomes in clinical settings. However, current methods for VO\textsubscript{2}max estimation, such as Cardiopulmonary Exercise Testing (CPET), require expensive equipment and the supervision of trained personnel, limiting accessibility for large-scale screening and prognostic purposes. Preliminary efforts have been made to create a more accessible method, such as the Cardiopulmonary Spot Jog Test (CPSJT). Unfortunately, these early attempts yielded high error margins, rendering them unsuitable for widespread use. In our study, we address these shortcomings by refining the CPSJT protocol to improve prediction accuracy. A key modification involved adjusting the CPSJT's cutoff point based on each participant's CPET-derived maximum heart rate (HR) rather than relying on a fixed percentage of theoretical max HR. This change provides a more individualized and accurate reflection of endurance. Our second crucial contribution is improved feature extraction and feature selection which include gender, body mass index, aerobic duration, and anaerobic duration. This targeted approach to feature extraction helps in streamlining the model to enhance prediction precision while minimizing the risk of overfitting. In a cohort of 44 participants from the Indian population, we assessed the performance of various machine learning models using these features. With Stratified 5-Fold Cross-Validation, the Root Mean Squared Error (RMSE) values were 5.78 for Linear Regression, 5.15 for Random Forest (RF), and 5.17 for Support Vector Regression (SVR). All models demonstrated strong test correlations and low RMSE values, underscoring their robust and reliable performance. This study highlights the potential of our enhanced CPSJT protocol in providing a more accessible and reliable method for non-invasive cardiovascular fitness evaluation, offering a promising alternative to traditional exercise testing methods.\\
\end{abstract}

\vspace*{-0.1\baselineskip}%
\renewcommand\IEEEkeywordsname{Keywords}

\section{Introduction}
Cardiovascular fitness is a fundamental aspect of overall health, playing a critical role in preventing and managing a wide range of chronic diseases, including heart disease, diabetes, and obesity \cite{r23}. As modern lifestyles increasingly lean towards sedentary behavior and poor dietary habits, the importance of maintaining cardiovascular health has never been more evident. One of the most widely recognized and clinically relevant measures of cardiovascular fitness is VO\textsubscript{2}max, which represents the maximum oxygen consumption during intense exercise \cite{r1}. VO\textsubscript{2}max is a robust indicator of aerobic endurance and overall cardiovascular health, making it a key metric not only in clinical settings but also in the optimization of athletic performance. The most accurate method for determining VO\textsubscript{2}max is through Cardiopulmonary Exercise Testing (CPET), which involves a graded exercise protocol, typically conducted on a treadmill or cycle ergometer, where the intensity is progressively increased until the subject reaches exhaustion \cite{r2}. Despite its accuracy, CPET is not without limitations. The test is resource-intensive, requiring specialized equipment, trained personnel, and a controlled environment. In light of these challenges, there is a critical need for alternative methods to estimate VO\textsubscript{2}max that are less resource-intensive, more accessible, and engaging for a broader population.

To address this need, a previous study by Mridula Badrinarayanan et al. \cite{r0} introduced a gamified protocol called the Cardiopulmonary Spot Jog Test (CPSJT) for the estimation of VO\textsubscript{2}max. This protocol aimed to provide a more accessible and engaging method for evaluating cardiovascular fitness. In their study, 22 participants underwent the CPSJT, and the estimated VO\textsubscript{2}max was compared with that obtained from CPET. They employed Random Forest (RF) and Support Vector Regression (SVR) models to predict VO\textsubscript{2}max using a wide range of features derived from heart rate (HR) and movement data captured during the CPSJT. However, the models exhibited considerable error margins, with the RF model yielding Root Mean Squared Error (RMSE) of 5.82 and the SVR model an even higher RMSE of 7.18 during K Fold cross-validation. These errors, coupled with the limited sample size, highlighted the challenges of overfitting and the need for further refinement.

In our study, we aim to address the shortcomings identified in the previous research by refining the CPSJT protocol and improving feature selection. Specifically, we expanded the participant cohort to 44 individuals from the Indian population and introduced a more streamlined approach to feature extraction. By focusing on a small set of key features—gender, Body Mass Index (BMI), aerobic duration, and anaerobic duration—we aimed to improve model robustness and accuracy while reducing overfitting. This targeted approach, in contrast to the previous study’s extensive feature set, enhances generalization, especially with simpler regression models \cite{r3}. Additionally, we adjusted the CPSJT termination criterion to be based on each participant's CPET-derived max HR, rather than a fixed percentage of the theoretical max HR, to provide a more accurate reflection of endurance and capacity. Various regression models were employed to assess the robustness of our approach. Furthermore, we re-evaluated RF and SVR models to compare their performance with the newly extracted features. Our findings suggest that the refined approach demonstrated substantial improvements in prediction accuracy. Regression models, when evaluated using Stratified K-Fold cross-validation, showed promising performance in terms of both robustness and accuracy, offering a compelling alternative to the more complex RF and SVR models, which also showed improved results. These improvements highlight the effectiveness of our refined methodology and underscore the potential of regression models, due to simplicity and lower susceptibility to overfitting \cite{r7}, making it well-suited for real-time applications.

Overall, our refined CPSJT protocol, coupled with strategic feature selection and the use of regression models, offers a practical, accessible, and reliable method for non-invasive cardiovascular fitness evaluation. This study highlights the potential of our approach as an alternative to traditional CPET, making cardiovascular fitness evaluation more engaging and widely applicable.

\begin{figure}[b]
\centerline{\includegraphics[width=0.33\textwidth]{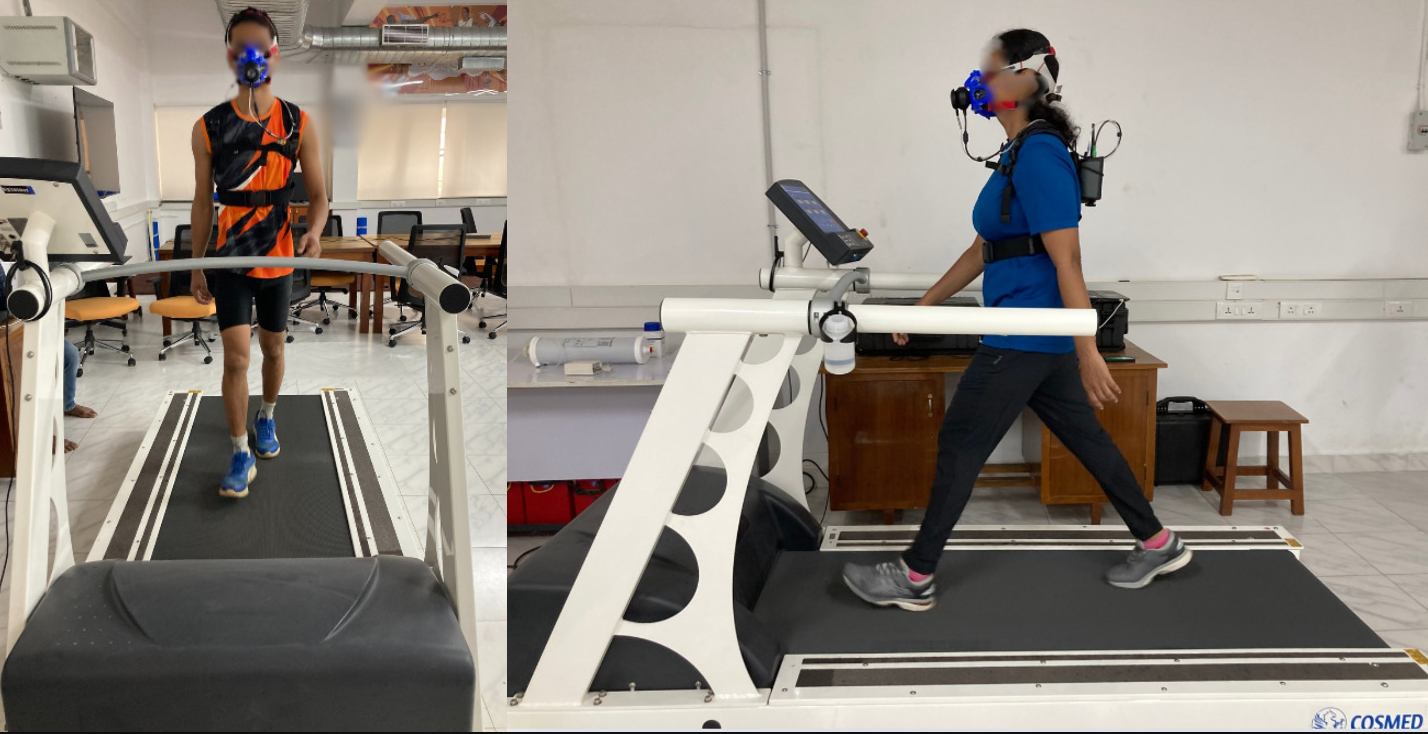}}
\caption{ Participants undergoing the Cardiopulmonary Exercise Testing (CPET) with the COSMED K5 device, which measures VO\textsubscript{2}max during the exercise protocol}
\label{fig-cosmed}
\end{figure}

\section{Methodology}
\subsection{Study Design and Data Collection}
The study involved a cohort of 44 healthy adults from the Indian population, comprising 30 males and 14 females. Although this distribution allowed for a comprehensive analysis of cardiovascular fitness metrics across genders, significant diversity in age groups was not achieved, with a majority of participants concentrated in the 18-30 age range. While the age distribution extended into the 30-45 range, many participants within this range were of similar ages, limiting the diversity of the sample. Specifically, the male participants were divided as follows: 19 in the 18-30 age group, and 11 in the 30-45 age group. For females, there were 8 participants in the 18-30 age group, and 6 in the 30-45 age group. Table \ref{table:demographics} provides a summary of the demographic parameters of the participants, including age, weight, height and BMI.

\begin{table}[ht]
\caption{Participants’ Demographic Characteristics}
    \centering
    \begin{tabular}{|l|c|c|c|c|}
        \hline
        \textbf{Parameters} & \multicolumn{2}{c|}{\textbf{Female}} & \multicolumn{2}{c|}
        {\textbf{Male}} \\
        \cline{2-5}
        & \textbf{Mean ± SD} & \textbf{Range} & \textbf{Mean ± SD} & \textbf{Range} \\
        \hline  
        Age (years) & 27.29 ± 7.74 & 19–44 & 27.6 ± 6.37 & 19–41 \\ 
        \hline
        Weight (kg) & 62.61 ± 11.1 & 51–88 & 71.92 ± 9.74 & 50–94 \\ 
        \hline
        Height (cm) & 160.01 ± 4 & 150–165 & 173.26 ± 6.44 & 161–186 \\ 
        \hline
        BMI (kg/m²) & 24.49 ± 4.36 & 19–33 & 23.92 ± 2.62 & 18–30 \\
        \hline
    \end{tabular}
    \raggedright 
    
    \footnotesize{\vspace{0.5em} SD: Standard Deviation}
    \label{table:demographics}
\end{table}

\textbf {Inclusion criteria}: Participants provided informed written consent as part of questionnaire process to confirm the absence of significant medical conditions, ensuring only healthy individuals from the Indian population were included.

This study was approved by Institutional Ethics Committee of Indian Institute of Technology Madras.\\

CPET was performed on the first day, and the CPSJT was conducted the following day. This sequencing was essential as the max HR recorded during CPET was used to set the termination criteria for the CPSJT. The data collection procedure followed the same general approach as our previous study \cite{r0} but incorporated significant improvements in the CPSJT protocol, which will be detailed further in this section.

\subsubsection{Cardiopulmonary Exercise Testing (CPET) Protocol}
On the first day, participants underwent CPET to establish a reliable ground truth for VO\textsubscript{2}max and maximum HR, as shown in Fig.~\ref{fig-cosmed}. The test utilized the COSMED K5 device \cite{r10} with the Balke-Ware treadmill protocol. To ensure that the max HR recorded was the participant's absolute maximum, participants were instructed to complete their last meal at least 2 hours before the test and ensure they were well-rested.
\begin{itemize}
    \item \textsl{Preparation and Warm-up:} Participants initially completed a 5-minute warm-up period involving light stretching and gentle movement. Following which, they were fitted with oxygen mask connected to the COSMED K5 device and Garmin HR sensor. Participants were briefed on the CPET protocol, including the various phases of the test, to ensure understanding.
    \item \textsl{Rest Period:} Participants rested for 2 minutes in a seated position before commencing the test.
    \item \textsl{Exercise Phase:} The test commenced with participants walking at a constant speed of 5 km/h. The treadmill inclination increased by 1\% every minute, continuing until the participant either completed the entire 25-minute test duration or reached voluntary exhaustion. Participants had the option to terminate the exercise phase early by raising their hand if they felt they had reached their maximum exertion. In such cases, they were immediately transitioned into the recovery phase.
    \item \textsl{Recovery Phase:} Following the completion of the exercise phase or early termination, participants entered a 2-minute recovery phase on the treadmill. During this phase, the treadmill speed was reduced to 3 km/h with no inclination. Post which the test concluded, and the equipment was removed.
\end{itemize}

\begin{figure}[htbp]
\centerline{\includegraphics[width=0.3\textwidth]{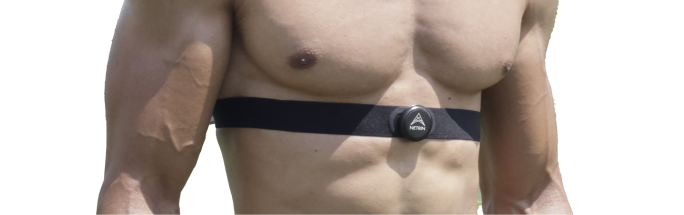}}
\caption{ Chest worn Netrin ECG \& motion sensor}
\label{fig-chest-sensor}
\end{figure}

\subsubsection{Cardiopulmonary Spot Jog Test (CPSJT) Protocol}
The following day, participants underwent CPSJT, an interactive digital game designed to predict VO\textsubscript{2}max that  involves spot jogging with a cadence increase every 30 seconds, continuing until they cannot maintain the required pace or reach their CPET max HR. This assessment used the Netrin Sensor, a chest strap device recording HR and acceleration data via single-lead ECG and a 3-axis accelerometer respectively as depicted in Fig.~\ref{fig-chest-sensor}, with real-time data transmission to the Netrin Enhance mobile application.
\begin{itemize}
    \item \textsl{Preparation and Setup}: Participants were fitted with the Netrin Sensor and connected to the Netrin Enhance app. They were briefed on the protocol, including details on the incremental exercise procedure and criteria for test termination.
    \item \textsl{Game Design and Execution:} Participants engaged in a 12-minute interactive game involving spot jogging. In this game, they controlled a character in an endless chase scenario, requiring them to maintain a specific cadence, as depicted in Fig.~\ref{fig-game-visual}. The game begins with a slow jog, with the cadence increasing every 30 seconds. Participants were guided by visual and audio cues to help them keep pace. The test concluded when participants failed to maintain the required movement index or if they reached their CPET max HR.
\end{itemize}

\begin{figure}[t]
\centerline{\includegraphics[width=0.33\textwidth]{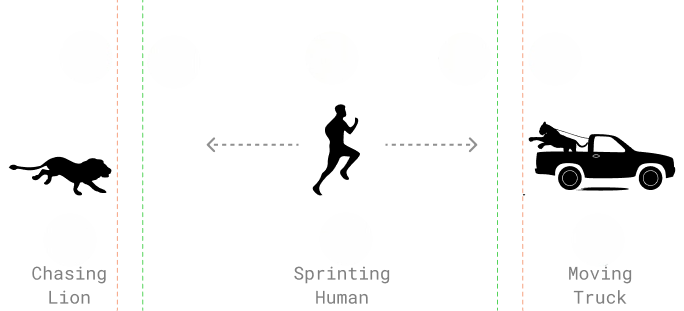}}
\caption{CPSJT interface wherein the player controls a sprinting character, moving horizontally based on their movement intensity. Visual markers include chasing lion and moving truck, representing the intensity range for each level. Player’s proximity to the lion indicates difficulty keeping up, while proximity to the truck shows overexertion.}
\label{fig-game-visual}
\end{figure}

\subsection{Modifications to CPSJT Protocol}
In this study, significant modifications were introduced to the CPSJT protocol, enhancing the accuracy and relevance of VO\textsubscript{2}max predictions as compared to the previous study \cite{r0}.
\begin{enumerate}
    \item \textsl{Adjustment of Test Termination Criteria:} Previously CPSJT was stopped when participants reached 90\% of their theoretical max HR. However, this criterion was found to be somewhat restrictive, as it often did not align with the participants' true exertion limits. To address this, the test is now terminated when participants reach their max HR that was recorded during CPET minus a 10 beat allowance. This adjustment was based on the observation that using the CPET max HR provided a more accurate representation of individual exertion limits. Most participants in the study felt they were at the verge of their maximum exertion with this revised approach, leading to more reliable assessment of their physical capacities.
    \item \textsl{Interpolation for Endured Duration:} 
    In some cases, participants did not reach their CPET max HR but were unable to generate sufficient movement through spot jogging to maintain the required cadence or movement index during higher levels of the test, due to which the test stopped. This difficulty occurred for a small subset of participants (6 out of the entire cohort) at higher intensity levels, wherein the sensor could not capture subtle rapid movements accurately as spot jogging limits the range of motion. To estimate the additional duration they could have sustained, first the difference between the participant's actual max HR during CPSJT and their CPET max HR was calculated. The time it took for the participant to reach the actual max HR from a lower value corresponding to this deviation provided an estimate of how long it would have taken them to reach their CPET max HR. To adjust for the fact that moving to higher HR generally happens faster, a factor of 75\% based on empirical observations was applied to this time estimate.
\end{enumerate}

\subsection{Statistical Analysis}
\subsubsection{Correlation and P-Value Overview}
The statistical analysis focused on exploring the relationships between VO\textsubscript{2}max and key metrics derived from CPSJT, namely aerobic duration, endured duration, and heart rate recovery (HRR) specifically for 30s, 60s and 120s periods. The correlation coefficients and p-values for these metrics with VO\textsubscript{2}max are summarized in Table \ref{table:correlation_pvalues}. 

Aerobic Duration emerges as a critical metric, demonstrating a stronger correlation with VO\textsubscript{2}max as compared to the Endured Duration. Aerobic Duration here is the time during which the participant's HR remains below 80\% of their CPET max HR throughout CPSJT. This period represents the duration when the participant is engaged in exercise at an intensity considered to be in the aerobic zone. In contrast, Endured Duration is the total time the participant was able to sustain the test, which continues until either the participant can no longer maintain the required cadence during the spot jog or they reach their CPET max HR.

\begin{table}[h!]
\caption{Correlation and P-Values for VO\textsubscript{2}max and Derived Metrics from CPSJT}
\centering
\begin{tabular}{|l|c|c|}
\hline
\textbf{Metric}            & \textbf{Pearson Correlation Coefficient} & \textbf{p-value} \\ \hline
\text{Aerobic Duration}  & 0.603100                         & 0.000015         \\ \hline
\text{Endured Duration}  & 0.544672                         & 0.000132         \\ \hline
\text{HRR 30s}           & 0.045742                         & 0.768116         \\ \hline
\text{HRR 60s}           & 0.096538                         & 0.533030         \\ \hline
\text{HRR 120s}          & 0.136856                         & 0.375701         \\ \hline
\end{tabular}
\label{table:correlation_pvalues}
\end{table}

Both these metrics derived from CPSJT capture different aspects of exercise performance, but Aerobic Duration, which specifically measures the time spent in aerobic exercise before transitioning to the anaerobic phase, is more closely aligned with VO\textsubscript{2}max. Since Aerobic Duration is a subset of Endured Duration, these variables are dependent, and including both could lead to redundancy and potential multicollinearity \cite{r14}. Therefore, Aerobic Duration is selected as a feature in our VO\textsubscript{2}max prediction model instead of Endured Duration. This choice highlights the significance of time spent within the aerobic threshold as a robust predictor of VO\textsubscript{2}max. The p-value of 0.000015 further reinforces the statistical significance of this finding.\\

\subsubsection{Heart Rate Recovery and Its Lack of Correlation with VO\textsubscript{2}max}
HRR identified by observing the steepest drop in HR within specified time windows - 30s, 60s and 120s from CPSJT showed weak and statistically insignificant correlations with VO\textsubscript{2}max as shown in Table \ref{table:correlation_pvalues}, despite HRR’s established role in cardiovascular fitness \cite{r12}. This discrepancy may stem from the passive recovery phase in CPSJT, which involves immediate seated rest period, unlike active recovery's gradual HR reduction through light exercise \cite{r13}. As a result, the HR drop observed during this recovery phase in CPSJT may not accurately reflect the participant's true recovery capacity, potentially masking the expected relationship with VO\textsubscript{2}max \cite{r22}. While HRR showed limited effectiveness in this study, incorporating a structured active recovery phase could enhance its predictive value in future models, emphasizing the importance of recovery protocol design in cardiovascular assessments.

\subsection{Feature Extraction}
This study utilized a limited set of features extracted from the CPSJT: gender, BMI, aerobic duration, and anaerobic duration, to predict VO\textsubscript{2}max. This contrasts with the previous study \cite{r0}, which relied on an extensive array of features, including HR time domain and frequency domain metrics, movement metrics, training load, recovery metrics, among others.

The choice of these selected features in our study was driven by their relevance to VO\textsubscript{2}max prediction and the goal of achieving model simplicity without sacrificing accuracy \cite{r9}. Gender and BMI were selected as they are well-established predictors of VO\textsubscript{2}max \cite{r18}. Gender differences are well-documented in aerobic capacity, with males generally exhibiting higher VO\textsubscript{2}max levels compared to females, partly due to differences in muscle mass and hemoglobin levels \cite{r15}. However, adding age as a feature did not enhance the model's performance due to the lack of age diversity in our cohort.

Aerobic duration, as previously discussed, was included for its strong correlation with VO\textsubscript{2}max. Anaerobic duration, defined as the period during CPSJT when the participant's HR exceeds 80\% of their CPET max HR, was also added as a feature. While aerobic duration is closely tied to VO\textsubscript{2}max, relying solely on it can overlook periods of high-intensity effort. For instance, some participants may achieve longer overall endured durations in CPSJT but spend significant time in the anaerobic zone, which could potentially be the reason they have lower VO\textsubscript{2}max values. Including anaerobic duration as a feature ensures that the model considers both sustained moderate activity and periods of intense effort, providing a more accurate prediction.

\subsection{VO\textsubscript{2}max Prediction Models}
In this study, we explored several Machine Learning (ML) models, with a preference for regression models due to their simplicity and robustness. Linear Regression, Ridge, Lasso, and Elastic Net were selected as the primary models. The model performance was evaluated using a refined set of significant but limited features, thereby avoiding the complexities associated with a broader array of features \cite{r8}. This targeted approach enhances not only model simplicity and accuracy but also makes these models more applicable in real-world settings. 

Additionally, we reevaluated RF and SVR, which had been employed in the previous study \cite{r0}. In the earlier work, these models were tested with a broader and more complex set of features, which  resulted in higher errors and increased risk of overfitting. 
The motivation behind revisiting RF and SVR was to assess whether a refined feature set, emphasizing fewer and more relevant predictors, could enhance VO\textsubscript{2}max prediction. This allowed for a more direct comparison with simpler regression models, ensuring that we explored both the complexity and simplicity trade-offs in model design without overfitting.
 
The flowchart in Fig.~\ref{fig-flowchart} provides an overview of the steps involved in the VO\textsubscript{2}max prediction process.

\begin{figure}[t]
\centerline{\includegraphics[width=0.5\textwidth]{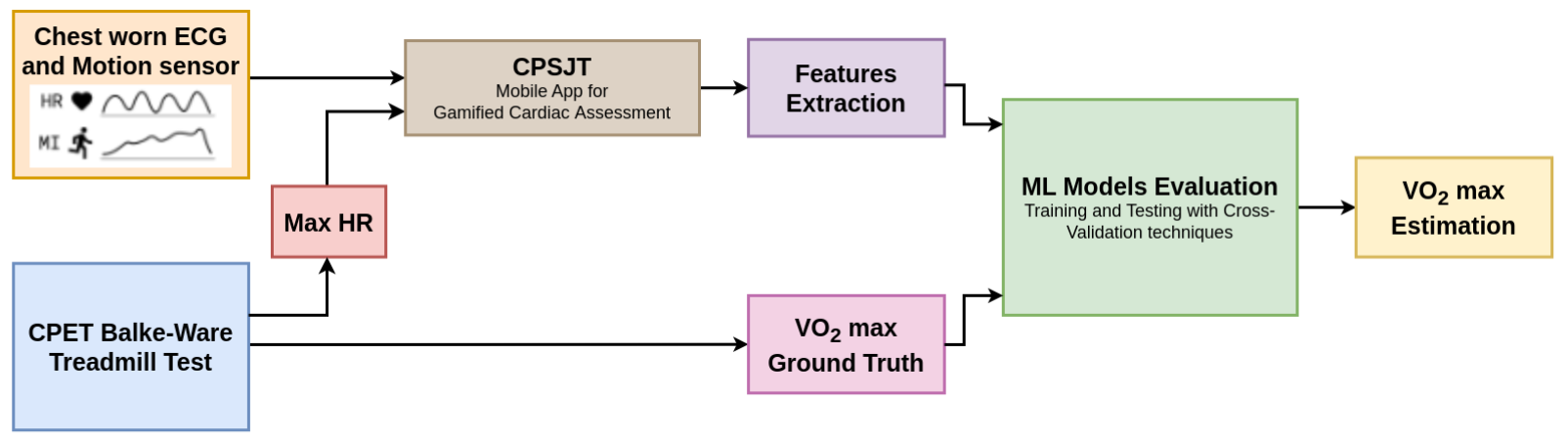}}
\caption{Flowchart for predicting VO\textsubscript{2}max from CPSJT using ML model}
\label{fig-flowchart}
\end{figure}

\section{Results and Discussion}
In this section, performance of the various ML models in predicting VO\textsubscript{2}max using metrics derived from the CPSJT are evaluated. We primarily focus on the errors associated with these models and the effectiveness of cross-validation to ensure robustness. All models were assessed using rigorous cross-validation techniques to ensure accuracy and generalizability. Linear regression, using Stratified 5-Fold cross-validation, achieved a mean RMSE of 5.78 with training and test correlations of 0.83 and 0.84, respectively. This model proved to be a strong alternative, maintaining simplicity without compromising accuracy. The more complex RF and SVR models also saw performance improvements, with RMSE values of 5.15 and 5.17, respectively, as compared to the previous study’s RF RMSE of 5.82 and SVR RMSE of 7.18

\begin{table*}[h!]
\caption{Comparison of Models using Stratified 5-Fold Cross-Validation and Leave-One-Out Cross Validation Techniques}
\centering
\begin{tabular}{|l|c|c|c|c|}
\hline
\textbf{Model} & \multicolumn{3}{c|}{\textbf{Stratified K-Fold}} & \textbf{LOOCV RMSE} \\ \cline{2-4}
               & \textbf{RMSE} & \textbf{Training Correlation} & \textbf{Test Correlation} &  \\ \hline
Linear Regression & 5.783177 & 0.834369 & 0.843257 & 6.184404 \\ \hline
Ridge             & 5.766069 & 0.834334 & 0.843002 & 6.154754 \\ \hline
Lasso             & 5.775528 & 0.834279 & 0.842088 & 6.170123 \\ \hline
Elastic Net       & 5.759368 & 0.834163 & 0.842156 & 6.129343 \\ \hline
RF                & 5.148647 & 0.952660 & 0.862395 & 5.689783 \\ \hline
SVR               & 5.170384 & 0.910955 & 0.874045 & 4.967588 \\ \hline
\end{tabular}
\label{table:combined_comparison}
\end{table*}

\subsection{Stratified K-Fold Cross-Validation}
To assess the performance and robustness of our models, we employed Stratified 5-fold Cross-Validation \cite{r20}. This approach ensured that the gender distribution within each fold was consistent with that of the original dataset. By maintaining a balanced male-to-female ratio across all folds, we minimized any potential gender-related biases that could affect the model’s performance. This stratification was essential for ensuring that the model's evaluation was representative and not skewed by imbalances in gender distribution.

The close alignment in RMSE and correlation values, as shown in  Table \ref{table:combined_comparison}, across the regression models - Linear, Ridge, Lasso, and Elastic Net emphasize the robustness and reliability of these simpler models, which are less prone to overfitting due to their inherent regularization techniques \cite{r5}. While RF and SVR showed competitive performance with lower RMSE and higher correlation values, the main objective was to assess the efficacy of a simpler feature set. The consistent results across regression models confirm that the selected features are effective for VO\textsubscript{2}max prediction, with RF and SVR used as benchmarks to validate that model simplification did not compromise accuracy. The minimal variation between training and test correlations across regression models indicate that they generalize well to new data. This consistency in performance reinforces the reliability of using these features across different model types, ensuring stable predictions across various subsets of data.

\subsection{Leave-One-Out Cross-Validation (LOOCV)}
To further validate the robustness of our models, we applied LOOCV. While LOOCV is particularly advantageous for small sample sizes, it remains a valuable technique even with our sample of 44 participants. This method involves using each data point as a test case while training the model on the remaining data points \cite{r19}. Results, as summarized in Table \ref{table:combined_comparison}, present the RMSE values across different models. While RMSE may not be exceptionally low, they are still sufficiently accurate for our predictive purposes across individual test cases. 

This use of LOOCV also helps in mitigating potential biases that might arise from arbitrary data splits, ensuring a more comprehensive assessment of the model’s predictive accuracy.

\begin{figure}[htbp]
\centerline{\includegraphics[width=0.43\textwidth]{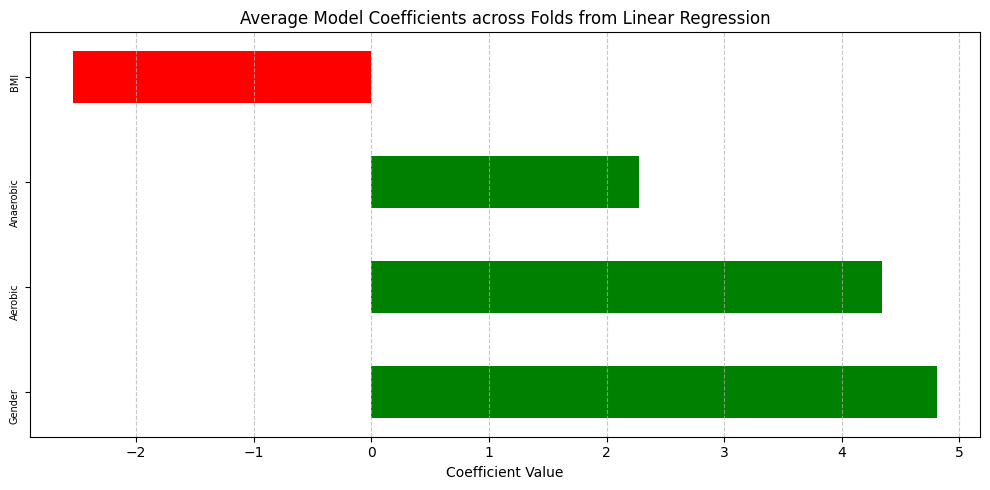}}
\caption{ Estimated coefficients for the Linear Regression model, corresponding to the features: gender, aerobic duration, anaerobic duration, and BMI, highlighting their influences in VO\textsubscript{2}max prediction. }
\label{fig-model-coeff}
\end{figure}

\subsection{Insights and Future Directions: Addressing Model Limitations and Enhancing Prediction Accuracy}
In the analysis, a notable observation emerged with one highly athletic male participant. Despite the generally improved accuracy of our models, this participant's predictions consistently exhibited substantial errors. Specifically, their actual VO\textsubscript{2}max was 61.9, while the prediction errors across models, calculated as the difference between actual and predicted VO\textsubscript{2}max, were as follows:  Linear Regression yielded an error of 12.39, Ridge 12.66, Lasso 12.86, Elastic Net 13.16, RF 11.24, and SVR 9.69. These errors indicate that the models consistently underestimated the VO\textsubscript{2}max for this participant. This participant was the sole individual in the cohort with a VO\textsubscript{2}max exceeding 55. The rest of the sample comprised individuals with generally lower VO\textsubscript{2}max levels, which constrained the model’s ability to generalize effectively to highly athletic individuals. This case underscores the need for a more diverse cohort that includes a larger number of athletes with high VO\textsubscript{2}max levels to enhance the model's predictive capabilities for such outliers.

Another notable insight from our analysis is the significant role of BMI as a predictor of VO\textsubscript{2}max, as evidenced by the model coefficients in Fig.~\ref{fig-model-coeff}. This highlights the influence of body composition on VO\textsubscript{2}max predictions. Although our study utilized BMI, it did not account for more granular body composition data. Future research should consider including such metrics to provide a deeper understanding of how different body types impact VO\textsubscript{2}max. This approach could improve prediction accuracy and offer better generalizability across various demographic groups.

\section{Conclusion}
This study examined the use of gamified cardiac assessment to estimate VO\textsubscript{2}max, focusing on the effectiveness of simplified models with a limited set of significant features to reduce the risk of overfitting while enhancing predictive accuracy. Through rigorous cross-validation techniques, we demonstrated the robustness and reliability of our predictive models. Regression models showed consistent performance, and more complex models like RF and SVR exhibited lower prediction errors. This indicates that the selected feature set is effective for VO\textsubscript{2}max prediction, and simpler regression models can be employed without significant loss of accuracy. This research marks a substantial step in translating advanced analytical methods into practical, accessible settings without significant accuracy trade-offs. The use of simple yet robust algorithms demonstrates their importance in real-world applications, making sophisticated health assessments more feasible and impactful in everyday contexts. Future research should focus on expanding the participant cohort to include a broader range of ages and fitness levels, as well as integrating detailed body composition metrics as features into the model. These steps will help to further refine the assessment and improve its applicability across different demographics, ensuring more accurate and individualized predictions.

\bibliographystyle{unsrt}
\bibliography{references}
\end{document}